A. KNIAZEVA (YANKEVICH), A. GOMAN

The Joint Institute of Mechanical Engineering of the NAS of Belarus, Minsk

E-mail: alena.kniazeva@tut.by

# DEFINITION OF A LOADING ZONE OF A PLANETARY PIN REDUCER ECCENTRIC

*Relationships are obtained for a definition of parameters which specify a location of an eccentric loading zone. This region is an expedient subject of an inquiry while performing works aimed at an optimization of dynamical and durability characteristics of a planetary pin reducer.*

**Introduction.** Planetary pin reducers (fig. 1) become more and more wildly applied in mechanical equipment. The reason for this is a number of advantages which planetary pin (cyclo) reducers obey as compared with involute reducers: wide range of reduction ratios, good specific mass and dimension characteristics, and high efficiency. That is why scientific researches and a pilot production of these perspective reducers are organized in the National academy of sciences of Belarus.

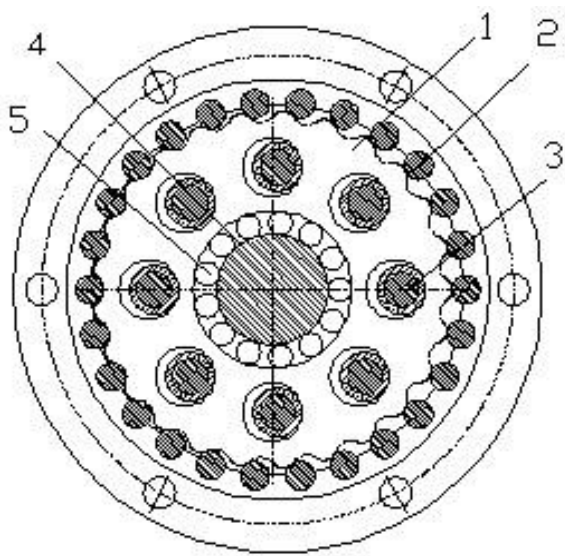

1 – satellite; 2 – pins; 3 – pins of parallel cranks mechanism;
4 – eccentric; 5 – rolling elements.

**Figure 1 – Cross-section of the planetary pin reducer**

Only a well-grounded engineering may provide excellent exploitation characteristics of the planetary pin reducer. Two problems may be noticed as main subjects of R&D activity in the field of planetary pin reducer engineering:
- decrease of dynamic loading of gear elements aimed on improvement of its vibroacoustic characteristics;
- increase of durability of the reducer.

Dynamical loads appear as results of manufacturing errors while producing of gear elements. That is why the solution of the first problem is achieved by providing of the prescribed accuracy of the satellites manufacturing. An original device for a gear-shaping machine is proposed in [1]. It allows to cut profiles of satellites using generating method, which is characterized as high-precision. Profiling of satellites by a contour creep feed method is also perspective. The effectiveness of this method is shown in a production of involute gear wheels [2, 3].

But it should be emphasized that decrease of dynamic loading may be achieved by an optimization of the eccentric unit construction. From this point of view constructional proposal [4] is very perspective. A method of compensation of errors in eccentric planetary pin reducer is proposed there. It implicates placing C-shape elastodeformed segments on the eccentric shaft. But authors did not take into account the direction of the load, which acts on the eccentric shaft. Segments are proposed to be placed orthogonally to the eccentric line [4]. But the load, which acts on the eccentric unit, is directed at angle to this line. That is why errors, which are results of inaccuracy while elements manufacturing and assembly, are compensated irregularly while the planetary pin reducer action. C-shape elastodeformed segments should be placed at the loading zone of eccentric, orthogonally to the direction of the radial force, acting on the unit.

As to the second problem, it is of the most actual for unit which durability is crucial for the durability of the whole reducer. The eccentric unit is such unit for the planetary pin reducer as follows from results presented in [5]. Generally accepted method of increase of durability is decrease of acting loads by means of rational selection of gear coupling parameters. Research [6] should be pointed from this point of view. An algorithm for determination of the geometrical parameter list is proposed in it to minimize dimensions of the planetary pin reducer. An additional optimization criterion is a decrease of a value of the load which acts on the eccentric unit.

Modernization of the eccentric unit construction is also a perspective direction of investigation aimed on decrease of the eccentric loading. It implies redistribution of loads between rolling bodies by means of placing elastodeformed segments (as described above) or correction of the eccentric profile.

So the determination of the eccentric loading zone is an actual problem. Its solution may be used in works aimed on decrease of dynamic loads which acts on the reducer elements so as increase the durability and improvement of loading conditions of the eccentric unit.

**Determination of the eccentric loading zone.** The loading zone of the eccentric is a part of its race on which the radial load is transmitted from the satellite through rolling bodies. It is bounded by an arc with angle equal to $180°$. This arc is orthogonal to the direction of the load acting on the eccentric unit (fig. 2). Described load is a geometrical sum of two components:

- load oriented along the eccentric line (the horizontal component $P_{12}^{H}$);
- load oriented orthogonal the eccentric line (the vertical component $P_{12}^{v}$).

The horizontal component $P_{12}^{H}$ of load changes on the load cycle in a coordinate system associated with the eccentric. The source of this change is the fact that values and the directions of loads in the coupling and the parallel cranks mechanism depend on the rotary angle of the input shaft. The vertical component remains constant in the same coordinate system. Its value may be determined from the relationship:

$$P_{12}^{v} = \frac{M_1}{2 \cdot e}, \qquad (1)$$

$M_1$ is a moment on the input shaft; e is a eccentricity of an epitrochoid, which describes the satellites profile.

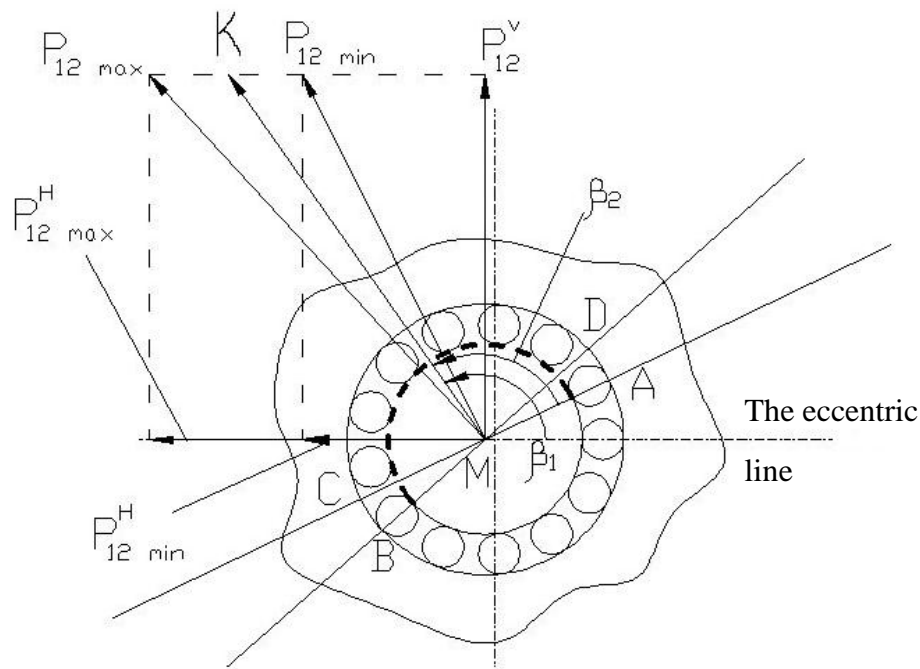

**Figure 2 — To the determination of the eccentric loading zone**

Radial load is taken by all rolling bodies located in a zone orthogonal to its direction and bounded with the angle $180^0$. That is why it may be concluded that in the case when the value of the radial load $\overline{P}_{12}$ become minimal $P_{12\,min}$ then it will act on the eccentric zone bounded with line AC at the figure 2. When its value is maximal $P_{12\,max}$ then BD is a zone bounding line. Pointed lines are orthogonal to directions of correspondent radial forces. The eccentric loading zone is bounded with external angle $\angle AMB$ (line AM is orthogonal to the direction of minimal force acting on the eccentric unit, BM – of maximal) as force direction changes in the sector between vectors $\overline{P}_{12\,min}$ и $\overline{P}_{12\,max}$. It is marked with dotted line at the figure 2. It is obvious that the arc specifying the eccentric loading zone may be determined by means of two angles in the polar coordinate system associated with the eccentric line. The first angle $\beta_1$ is directional angle. It is specify the middle of the eccentric loading zone in respect to polar axis MO. The second angle $\beta_2$ is bounding one. It is the angle between bounding vector $\overline{MA}$ (which is corresponding to the direction of the minimal load $\overline{P}_{12\,min}$) and middle vector $\overline{MK}$. Values of these angles may be determined on the base of next relationships:

$$\beta_1 = \pi - \frac{1}{2}\left(\arcsin\left(\frac{P_{12}^v}{P_{12}^{max}}\right) + \arcsin\left(\frac{P_{12}^v}{P_{12}^{min}}\right)\right) =$$
$$= \pi - \frac{1}{2}\cdot\left(\arcsin\left(\frac{M_1}{\sqrt{4\cdot e^2\cdot \left(P_{12}^{H\,min}\right)^2 + M_1^2}}\right) + \arcsin\left(\frac{M_1}{\sqrt{4\cdot e^2\cdot \left(P_{12}^{H\,max}\right)^2 + M_1^2}}\right)\right); \quad (2)$$

$$\beta_2 = \beta_1 - \frac{\pi}{2} + \arcsin\left(\frac{P_{12}^v}{P_{12}^{min}}\right) = \frac{\pi}{2} + \frac{1}{2}\cdot\left(\arcsin\left(\frac{P_{12}^v}{P_{12}^{min}}\right) - \arcsin\left(\frac{P_{12}^v}{P_{12}^{max}}\right)\right) =$$
$$= \frac{\pi}{2} + \frac{1}{2}\cdot\left(\arcsin\left(\frac{M_1}{\sqrt{4\cdot e^2\cdot \left(P_{12}^{H\,min}\right)^2 + M_1^2}}\right) - \arcsin\left(\frac{M_1}{\sqrt{4\cdot e^2\cdot \left(P_{12}^{H\,max}\right)^2 + M_1^2}}\right)\right), \quad (3)$$

$P_{12}^{max}$ and $P_{12}^{min}$ are accordingly maximal and minimal values of force, which acts on the eccentric unit on the loading cycle; $P_{12}^{H\,max}$ and $P_{12}^{H\,min}$ are their horizontal components.

The horizontal component $P_{12}^H$ of the force, which acts on the eccentric unit, is calculated according to [7] on the base of the next relationship:

$$P_{12}^H = \sum \Pr\nolimits_{ecc\,line} \overline{N}_{4i} - \sum N_{3j}, \quad (4)$$

$\Pr_{ecc\,line} \overline{N}_{4i}$ is a projection of a force in the "satellite- i-pin" coupling $\overline{N}_{4i}$ on the direction of the eccentric line; $N_{3j}$ is a force, which acts on the satellite from the crankshaft under number j.

As follows from results, obtained in [7], a function, which describes change of the sum $\sum N_{3j}$ on the load cycle, may be specified as follows:

$$\sum N_{3j} = \frac{M_1 z_2}{2r_3} \cdot \frac{\sum_j \sin\left(\frac{2\cdot\pi\cdot j}{z_3} - \varphi\cdot\left(1+\frac{1}{z_2}\right)\right)}{\sum_j \sin^2\left(\frac{2\cdot\pi\cdot j}{z_3} - \varphi\cdot\left(1+\frac{1}{z_2}\right)\right)}, \qquad (5)$$

$z_2$ is a number of satellite teeth; $r_3$ is a radius of a circle, which pass through all of crankshafts centers; $z_3$ is a number of crankshafts; $\varphi$ is a parameter, which describes an angle of eccentric rotation relatively to its initial position $\varphi \in [0, 2\pi]$.

A series of numerical experiments, performed using Mathematica 5.0, showed the validity of the next approximation:

$$\sum_j \sin^2\left(\frac{2\cdot\pi\cdot j}{z_3} - \varphi\cdot\left(1+\frac{1}{z_2}\right)\right) \approx \frac{z_3}{4}. \qquad (6)$$

Let's substitute (6) into (5). Then we obtain that the sum of projection of forces which act on the satellite from parallel crankshaft mechanism on the load cycle may be determined as follows

$$\sum N_{3j} = \frac{2M_1 z_2}{r_3 z_3} \cdot \sum_j \sin\left(\frac{2\cdot\pi\cdot j}{z_3} - \varphi\cdot\left(1+\frac{1}{z_2}\right)\right) = \frac{2M_1 z_2}{r_3 z_3} \cdot S. \qquad (7)$$

The function $\sum N_{3j}$ is periodic. It's periodicity is conditioned by the multiplier $S$. The plot of this multiplier is shown at the figure 3 against to the eccentric rotation angle. It is obvious that angular pitch of the $S$-function is equal to $2\pi z_2 / z_3(z_2 + 1)$.

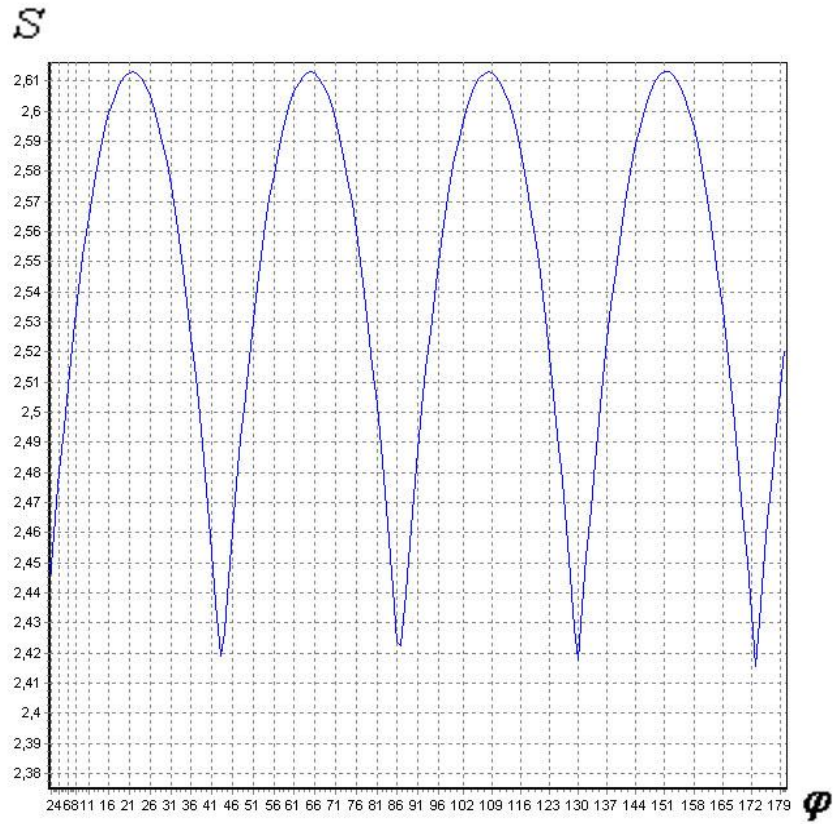

**Figure 3** — Change of $S = \sum_j \sin\left(\dfrac{2\cdot\pi\cdot j}{z_3} - \varphi\cdot\left(1 + \dfrac{1}{z_2}\right)\right)$ on the load cycle

An approximation of the sum S from the right side of equation (7) was founded on the base of a series of numerical experiments and analysis of the graph presented at the figure 7. Then a relationship was obtained for the calculation of the sum $\sum N_{3j}$ of projections on the eccentric line of forces, which act on the satellite from the parallel crankshaft mechanism:

$$\sum N_{3j} = \frac{2\cdot M_1 \cdot z_2}{r_3 \cdot z_3} \cdot \left[ P_1 + P_2 \cdot \left|\sin\left(\frac{(z_2+1)\cdot z_3}{2\cdot z_2}\cdot \varphi\right)\right| \right], \qquad (8)$$

$P_1$, $P_2$ are functions obtained on the base of interpolation of the sum $\sum_j \sin\left(\dfrac{2\cdot\pi\cdot j}{z_3} - \varphi\cdot\left(1 + \dfrac{1}{z_2}\right)\right)$ values, performed using Mathematica 5.0:

$$P_1 = -1{,}215 + 0{,}725 \cdot \sqrt{z_3} + 0{,}1863 \cdot z_3 + 0{,}00141 \cdot z_3^2 ;$$
$$P_2 = 1{,}8581 - 1{,}1127 \cdot \sqrt{z_3} + 0{,}2031 \cdot z_3 - 0{,}002175 \cdot z_3^2 .$$

A calculation was performed for the planetary pin motor-reducer МПЦЦ 82-G25, which characterized with the next parameters:

- the moment on the input shaft $M_1 = 5{,}94$ N·m;
- the radius of the pin gear ring $r = 53{,}5$ mm;
- the number of satellite teeth $z_2 = 25$;
- the number of crankshafts $z_3 = 8$;
- the radius of the circle, which pass through all of crankshafts centers $r_3 = 35$ мм.

Performed calculations showed that a relative error of the proposed approximation of the component $\sum N_{3j}$ is not more than 0,2%.

Approximation for the sum $\sum \text{Pr}_{\text{ecc line}} \overline{N}_{4i}$ was obtained in the same way:

$$\sum \text{Pr}_{\text{ecc line}} \overline{N}_{4i} = -\frac{M_1}{2 \cdot e} \cdot \left( A_1 - A_2 \cdot \left| \sin\left( \frac{z_2 + 1}{2} \cdot \varphi \right) \right| \right), \qquad (9)$$

$A_1$, $A_2$ are interpolation functions which are obtained using computer program Mathematica 5.0:

$$A_1 = 0{,}276 - 0{,}28 \cdot k + 0{,}697 \cdot k^2 - 0{,}0049 \cdot z + 0{,}000073 \cdot z_2^2;$$
$$A_2 = 0{,}057 + 0{,}000056 \cdot k - 0{,}0431 \cdot k^3 + 0{,}14 \cdot k^5 - 0{,}0019 \cdot z_2;$$

k is a epitrochoid shortening coefficient, $k = \dfrac{e \cdot (z_2 + 1)}{r}$;

r is the radius of the pin gear ring;

e is the eccentricity of the epitrochoid.

A calculation was performed for the planetary pin motor-reducer МПЦЦ 82-G25. It showed that the relative error of the proposed $\sum \text{Pr}_{\text{ecc line}} \overline{N}_{4i}$ approximation is not higher than 1%.

By means of substituting (8) and (9) into (4) we obtain the function which describes dependence of the horizontal component value of the force, which acts on the eccentric unit, from the eccentric rotation angle. But we cannot calculate extremes of the function obtained. When we find the derivative of this function and put it equal to zero we obtain transcendental equation. And it's impossible to determine its roots explicitly. That is why it is reasonable to use graphic method to solve this problem. Graphs of dependence of functions $P_{12}$, $\sum Pr_{ecc\ line}\ \overline{N}_{4i}$ and $P_{12}^H$ from eccentric rotation angles are shown at the figure 4 for the planetary pin motor reducer МПЦЦ 82-G25.

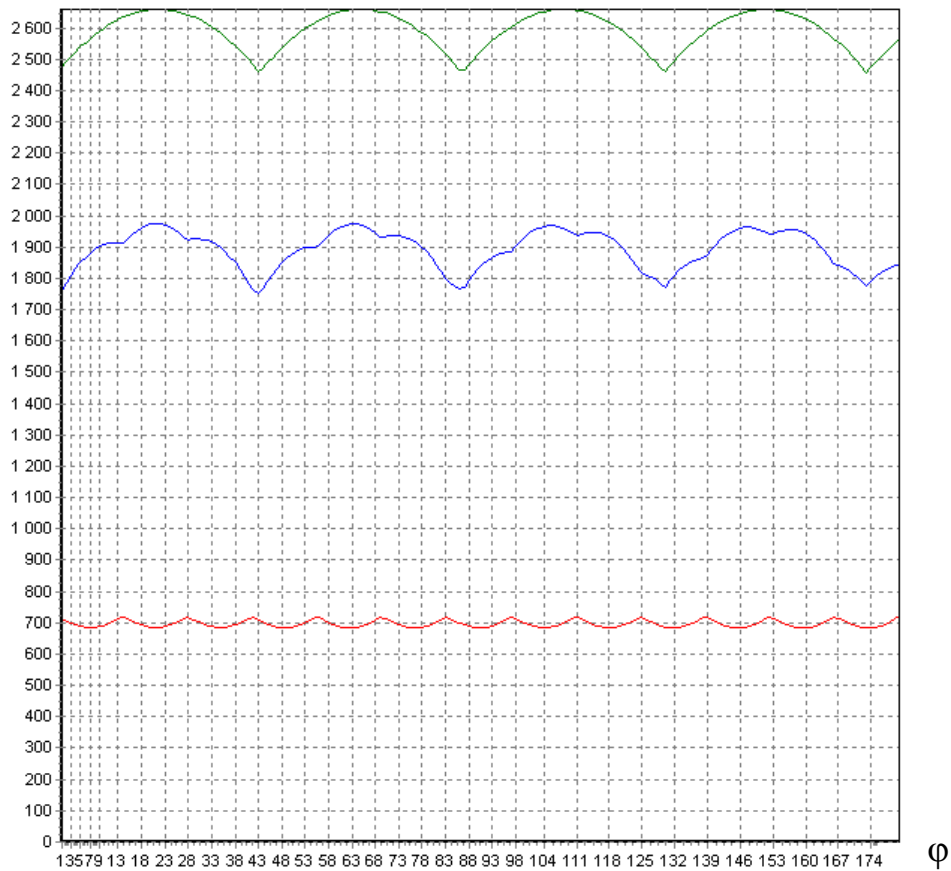

▬ — the sum of projections of forces, which act on the satellite from the parallel crankshafts mechanism ($\sum N_{3j}$), N;

▬ — the horizontal component of the force, which acts on the eccentric unit ($P_{12}^H$), N;

▬ — the sum of projections of forces which act from pins ($\sum Pr\ N_{4i}$), N.

**Figure 4 — Changing of constitutive sums of the horizontal component of the force, which acts on the eccentric unit on the load cycle**

It is obvious that formation of extremes of the function $P_{12}^H$ is mainly influenced by the component $\sum N_{3j}$. That is why it can be concluded that $P_{12}^H$ is minimal when the rotational angle is equal to zero. And it takes on the maximum value when the eccentric is rotated on the angle $\varphi = \dfrac{\pi \cdot z_2}{z_3 \cdot (z_2 + 1)}$ with respect to the initial position.

Then it is obtained from relationships (4), (8) and (9):

$$P_{12}^{H\,min} = \frac{2 \cdot M_1 \cdot z_2 \cdot P_1}{r_3 \cdot z_3} - \frac{M_1 \cdot A_1}{2e};$$
$$P_{12}^{H\,max} = \frac{2 \cdot M_1 \cdot z_2}{r_3 \cdot z_3}(P_1 + P_2) - \frac{M_1}{2e}\left(A_1 - A_2 \cdot \left|\sin\left(\frac{\pi \cdot z_2}{2 \cdot z_3}\right)\right|\right). \qquad (10)$$

Angles, bounding the eccentric loading zone, can be obtained after the substitution of $P_{12}^{H\,min}$ и $P_{12}^{H\,max}$ (which are determined by (10)) into (2) and (3).

Performed calculation showed that the loading zone of the eccentric of the planetary pin reducer МПЦЦ 82-G25 is determined by angles $\beta_1 = 131°$, $\beta_2 = 92°$. Namely this zone should be the object of researches aimed on the decrease of dynamic loading and increase of the durability of the eccentric of the planetary pin reducer.

**Recommendations for the decrease of dynamic loading of gear elements and increase of the eccentric unit durability.**

*Decrease of dynamic loading* may be achieved by means of placing of C-shape elastodeformed segments on the eccentric shaft similarly to the solution proposed in [4]. Elastodeformed segment should be placed accurate in the loading zone. Symmetry line should be directed along the ray which corresponds to the angle $\beta_1$ in polar system, connected to the eccentric. Coupling errors in the planetary pion reducer will be compensated uniformly in such a case.

*Increase of the durability.* A breakdown of the planetary pin reducer begins from a pitting of the eccentric. It takes place in a zone (loading zone) where it contacts with rolling bodies under action of the radial load, which is transmitted from the satellite. That is why hardening of the eccentric loading zone is an effective method to increase the durability both eccentric unit and whole planetary pin reducer.

*Increase of the durability by means of decrease of the elements loading.* Increase of the eccentric unit durability may be achieved through the decrease of the loading level for unit's elements. A half of all rolling bodies transmit the load in a typical eccentric unit. The maximal load is transmitted through the body, whose center is the closest to the direction vector of the radial load $\overline{P}_{12}$, which acts on the unit. Rolling bodies, which are situated near the border of the eccentric loading zone, doesn't take part in an actual load transmission. That is why an eccentric loading zone correction may be an effective method to increase the durability of the eccentric unit. The essence of this correction is to bring all rolling bodies, which are in the loading zone, to transmit approximately equal parts of the radial load.

**Conclusion.** The load, which acts on the eccentric unit on the load cycle, is transmitted to the finite eccentric zone. This zone may be determined in a polar system, associated with the eccentric. Relationships are proposed in this article for the calculation of directional and bounding angles, which describe the position of the eccentric loading zone. Values of these angles depend on value of the rotation moment on the reducer input shaft and parameters of the epitrochoid, which describes the satellite profile. Performed calculation showed that the loading zone of the eccentric of the planetary pin reducer МПЦЦ 82-G25 is determined by the directional angle $\beta_1 = 131°$ and the bounding angle $\beta_2 = 92°$. Knowledge about the eccentric loading zone is extremely important. It can be used while development and improvement of methods for the decrease of dynamic loading of gear elements and increase of the eccentric unit durability.